\documentclass[%
fleqn,%
final,%
]{an}
\usepackage{graphicx}
\usepackage{times}
\usepackage{amssymb}
\usepackage{amsmath}
\begin{document}

\Pagespan{1}{}
\Yearpublication{2013}%
\Yearsubmission{2013}%
\Month{4}%
\Volume{334}%
\Issue{8}%
\DOI{This.is/not.aDOI}%

\title{Metallicity field and selection effects in spatial distribution\\[1.5pt] of the
 Galactic globular cluster system}

\author{I.I. Nikiforov\thanks{Corresponding author:
{nii@astro.spbu.ru}} \and O.V. Smirnova }

\authorrunning{I.I. Nikiforov \& O.V. Smirnova}
\titlerunning{Metallicity field and selection effects for the Galactic globular
cluster system}

\institute{Sobolev Astronomical Institute, Saint~Petersburg State University,
Universitetskij Prospekt 28, Staryj Peterhof,\\ Saint~Petersburg 198504, Russia}

\received{2013}
\accepted{2013}
\publonline{2013}



\abstract{%
The prospects for using the present-day data on metallicity of globular clusters (GCs) of
the Galaxy to put constraints on the distance to the Galactic center, $R_0$, are
considered. We have found that the GCs of the metal-rich and metal-poor subsystems
separately form a bar-like structure in metallicity maps whose parameters are very close
to those for the Galactic bar. The results indicate the existence of a bar component
within both the metal-rich and metal-poor subsystems of GCs. The bar GCs could have formed
within the already existing Galactic bar or could have later been locked in resonance with the bar.
We conclude that substantial constraints on the $R_0$ value can be obtained only with
non-axisymmetric models for the space distribution  of GC metallicities with the allowance
for the subdivision of GCs into subsystems. We found evidence for a bar
extinction component that causes the observational incompleteness of GCs in the far side of
the Galactic bar and in the ``post-central'' region. This selection effect should be taken
into account when determining $R_0$ from the spatial distribution of GCs. }

\maketitle


\section{Introduction}

The study  of the spatial distribution of metallicity for the Galactic system of
globular clusters (GCs) helps to reveal the properties of this system, which are of importance
for our understanding of the formation and evolution processes of the system and of the
whole Galaxy. In particular, the analysis of this distribution, along with that of
kinematics, made it possible to establish the division of GCs into two subsystems:
metal-rich disk GCs and metal-poor halo GCs (Zinn~1985). More recently, 
halo GCs, in turn, were shown to be subdivided into at least two groups based on their
horizontal branch morphology, kinematics, and other parameters (Zinn~1993; Da Costa \&
Armandroff~1995; Borkova \& Marsakov 2000; see also Bica et al.~2006 and references
therein). In addition, three subsystems were identified among the metal-rich GCs (Burkert
\& Smith 1997).

Another problem that is discussed extensively is that of the existence of metallicity gradients
in the GC (sub)system(s) (e.g., Zinn~1985; Alfaro, Cabrera-Ca\~{n}o \& Delgado~1993;
Borkova \& Marsakov 2000; van den Bergh 2011). This problem, in addition to its importance in
itself, is  associated with the problem of the determination of the distance to the center of
the Galaxy, $R_0$. For one thing, an investigation of the spatial metallicity distribution
(as well as merely the spatial distribution) of GCs requires the 
distance scale for GCs to be compatible with the adopted $R_0$. For another, the existence of a
{\em radial} gradient, more generally the dependence of [Fe/H] on the distance of the GC
from the Galactic  axis, $R$, can, in principle, impose constraints on the value of~$R_0$.

Surdin (1980) suggested a method of estimating $R_0$ based on the [Fe/H]-$R$ relationship
assuming that the  GC distribution in the coordinate--metallicity $(X, Y,
Z, {\rm [m/H]})$ space is axisymmetric.  However, the $R_0$ values found by Surdin (1980) from {\em all\/} GCs
without subdividing them into subsystems, $R_0=9.9\pm 0.3$~kpc and $R_0=10.3\pm 0.6$~kpc for two
catalogues of GCs, now appear to be overestimated. This cannot be explained by the
evolution of the distance-scale calibration: rescaling to the current calibration 
\begin{equation}
\label{H10calG}
    M_V(\mathrm{HB})=0.165\,\rm{[Fe/H]}+0.86,
\end{equation}
based on the most direct distance measurements {\em within the Milky Way} (see the
2010 edition of the Harris (1996) catalogue), yields $R_0=10.8\pm 0.3$~kpc and
$R_0=10.1\pm 0.6$~kpc, respectively. Although the revised $R_0$ estimate found by applying this method
to the original version of the Harris (1996) catalogue is $8.6\pm 1.0$~kpc (Surdin 1999),
rescaling it to  calibration~(\ref{H10calG}) gives a larger value of $9.0\pm
1.0$~kpc.


\begin{table*}
\caption{Average  $R_0$ values found using different  groups of estimates.}%
\label{tab:R0average}
\begin{tabular}{lcccc}
\hline\noalign{\smallskip}
Group of Estimates & $N_{\mathrm{est}}$ &$N_{\mathrm{pap}}$ &$N_{\mathrm{hg}}$ & 
$\langle R_0\rangle$ \\ 
&&&&(kpc)\\

\hline\noalign{\smallskip}
Based on GCs, all methods$\,^\mathrm{a}$		& 19 & 13 & \enspace3 &	$7.63\pm 0.38\,^\mathrm{b}$\\
Based on GCs, all spatial methods		& 16 & 10 &\enspace 1 &	$7.42\pm 0.23\,^\mathrm{b}$\\
Based on GCs, Shapley's method and related spatial methods	
					& 12 & \enspace8  &\enspace 1 & $7.36\pm 0.24\,^\mathrm{b}$\\
\hline\noalign{\smallskip}
All estimates				& 74 & 60 & 18 & $7.91\pm 0.15\phantom{\,^\mathrm{b}}$\\
\hline
\noalign{\smallskip}
\multicolumn{5}{l}{$^\mathrm{a}$ Including Surdin's method with the estimate by Surdin (1999).}\\

\multicolumn{5}{l}{$^\mathrm{b}$ The quoted error does not include the systematic
uncertainty of the adopted distance scale.}
\end{tabular}
\end{table*}

Let us now compare the estimates obtained using Surdin's method with other $R_0$ estimates
based on GC data and with the
current best values of $R_0$. Table~\ref{tab:R0average} lists the average $R_0$ values
derived by applying the same procedure as used by Nikiforov (2004) to a selection of groups of $R_0$
estimates. We use an updated version of Nikiforov's (2004) sample of $R_0$
estimates published since 1974. In the case of GC-based estimates, only the result by Bica
et al. (2006), $R_0=7.2\pm 0.3$~kpc, was added.  All GC-based estimates are rescaled
according to calibration~(\ref{H10calG}). In the table, $N_{\mathrm{est}}$ is the number
of $R_0$ estimates; $N_{\mathrm{pap}}$, the number of papers, and $N_{\mathrm{hg}}$,
the number of homogeneous groups of $R_0$ estimates, i.e., based on the same class of
methods, the same class of reference distances, and the same type of reference objects 
(see Nikiforov 2004). 
The uncertainty of the average value, $\langle R_0\rangle$, listed in Table~\ref{tab:R0average} 
for groups of GC-based estimates reflects the statistical uncertainty and the systematic uncertainty of
the method used to derive the $R_0$ estimate from the adopted reference distances (see Nikiforov 2004), 
but does not include the
systematic uncertainty of the adopted distance scale, because the latter is the same for
all GCs' groups. The uncertainty in value of $\langle R_0\rangle$ derived from all
estimates is a combination of all errors.

The bottom entry in Table~\ref{tab:R0average} shows that the $R_0$ estimates based on the
radial metallicity gradient are essentially greater than the mean value of $R_0=7.9\pm
0.2$~kpc averaged over all methods and objects -- the so called ``best value'' for $R_0$; cf. the best
estimates of $R_0=(8.15$--$8.25)\pm (0.14$--$0.20)$~kpc by Genzel, Eisenhauer \&
Gillessen (2010) deduced from 11 recent (2006--2009) $R_0$ estimates. However, the
discrepancy between the results obtained by Surdin's method and those obtained using other 
GC-based techniques is, on
average, even greater (Table~\ref{tab:R0average}), although all these $R_0$ estimates were
rescaled to the same calibration. 

Such discrepancies may be due to incorrect assumptions adopted in Surdin's method.
In particular, the method does not allow for the fact that the GC populations 
consist of the metal-rich and metal-poor subsystems, that is, it
assumes, in fact, the existence of a {\em smooth} radial metal\-licity gradient for the
entire system of GCs; however, now we see that this appears not to be the case, rather the
gradient is like to a stair-step (e.g., Borkova \& Marsakov 2000; van den Bergh 2011). If
so, the efficiency and systematics of Surdin's method depend on the existence of
radial gradient in the metal-rich and/or metal-poor subsystems separately. However, radial
gradients are found to be insignificant for the most of GCs' subsystems identified
(e.g., Borkova \& Marsakov 2000), at least for a fixed value of $R_0$. If there are no
radial gradients within the metal-rich and metal-poor subsystems, Surdin's method reduces
to the determination of the centroid of distribution of metal-rich GCs, i.e., becomes akin
to Shapley's method and related ones (see, e.g., Reid 1993; Nikiforov 2004). If so, then
it is not clear why $R_0$ estimates found by these two approaches are so different
(Table~\ref{tab:R0average})? It is only clear that in this case the allowance
for selection effects in the distribution of GCs caused by extinction becomes as important
for Surdin's method as it is for  Shepley's method. In both methods, simulations were performed
to estimate the bias (Surdin 1999; e.g., Racine \& Harris 1989), but the results of
such modelling depend on the assumptions concerning the extinction law. This may be an
additional source of systematic error in both methods.

The starting point for  this work was to clarify, based on the current knowledge of the
Galactic GC system, whether the present-day GC metallicity data can impose (significant)
constraints on $R_0$. Pursuing this goal has sent us to evaluate the uncertainty of
new metallicity data and study the details of the GC distribution. In this paper we present
some of the results obtained.

\section{Data on globular clusters}

Our GC data is the 2010 December version (hereafter H10) of the
Catalog of Parameters for Globular Clusters in the Milky Way by Harris (1996), ${N_{{\rm
tot}}=157}$. The catalog presents the distance estimates (for all GCs) calculated
using the calibration
\begin{equation}
\label{H10cal}
    M_V(\rm{HB})=0.16\,\mathrm{[Fe/H]}+0.84,
\end{equation}
which is based on the most direct distance measurements for objects in the Milky
Way and on the distances to GCs {\em in M31} found with an adopted fiducial distance for
M31. Thus this calibration is to some extent secondary compared to
calibration~(\ref{H10calG}). The latter is only slightly fainter (0.01--0.02~mag) than 
calibration~(\ref{H10cal}).

For 152 GCs, the new list provides  ${\rm [Fe/H]}$ values that are on a {\em new
metallicity scale} based on high dispersion spectroscopy (Carretta et al. 2009). This
represents a fundamental change from the older metallicity scale by Zinn \& West (1984)
used in the 2003 version (hereafter H03) and other previous editions of the catalog. The
H10 list also provides weights for ${\rm [Fe/H]}$, $p$, which are essentially equal to 
the number of independent [Fe/H] measurements averaged for each GC.


\begin{figure*}
\center%
\vspace*{-36pt}
\hspace*{-50pt}%
\includegraphics[angle=0,width=0.8\textwidth,clip]{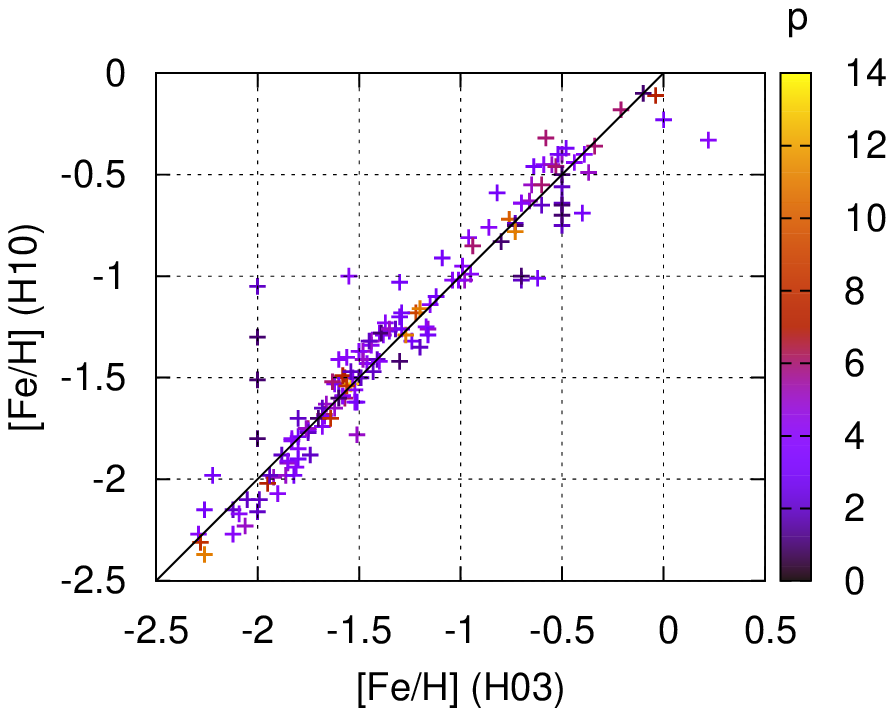}%
\raisebox{8pt}{%
\hspace*{-43pt}%
\includegraphics[angle=0,width=0.39\textwidth,clip]{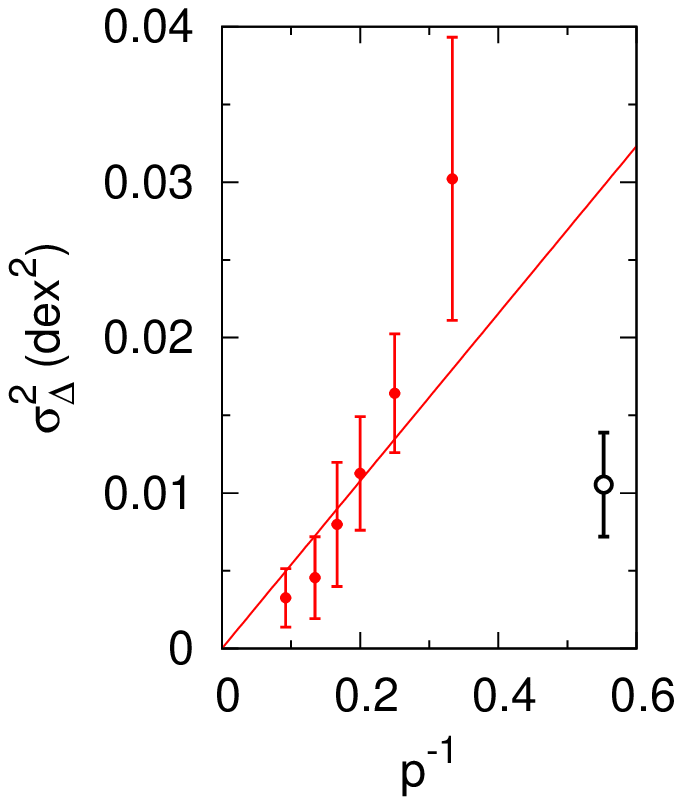}%
}%
\vskip-4mm
\caption%
{\emph{Left}: Comparison of metallicities [Fe/H] from the 2010 (H10) and 2003 (H03) versions of
the Harris (1996) catalogue. The solid line corresponds to equal metallicities. The palette represents
the H10 metallicity weights, $p=1,\,\ldots,\,13$. \emph{Right}: The variance of differences
between the H10 and H03 metallicities versus the inverse of H10 metallicity weight,
$p^{-1}$. The line is the best-fit regression~(\ref{varianceFeH}) to all data
points except that obtained for $p=1,2$ (the open circle).}%
\label{fig:metH10-03}
\end{figure*}


Figure~\ref{fig:metH10-03} (left panel) gives  an idea of the systematic differences between the two
metallicity scales, i.e., between the metallicities ${\rm [Fe/H]}_{10}$ and ${\rm
[Fe/H]}_{03}$ in the two versions of the catalogue. The vertical chains of data points with
large residuals relative to the line of equality at ${\rm [Fe/H]}_{03}=-2.00$ and $-0.50$
suggest that some of these ${\rm [Fe/H]}$ values in H03 were adopted rather than
estimated. To evaluate the dependence of uncertainties of ${\rm [Fe/H]}_{10}$ on weights $p$, we
computed the variances $\sigma^2_{\Delta}$ of the differences 
\begin{equation}
\label{Delta10-03}
\Delta= {\rm [Fe/H]}_{10}-{\rm [Fe/H]}_{03} 
\end{equation}
for common GCs both for  $p$-based bins and for all 125 GCs in common. In doing so, we
excluded the GCs with ${\rm [Fe/H]}_{03}=-2.00,-0.50$ and ${\rm [Fe/H]}_{10}=-1.00$ as
``blunders'' (see Fig.~\ref{fig:metH10-03}, left panel) as well as GCs with unchanged
estimates of ${\rm [Fe/H]}$ (${\Delta=0.00}$). It appears from Fig.~\ref{fig:metH10-03} (right
panel) that the variances $\sigma^2_{\Delta}$ and weights are  very highly correlated,
except for one data point in the lowest-weight bin, $p=1,2$, because of an overlap between the
H03 and H10 source lists for metallicities in this case. Without this point, the linear
correlation coefficient for the  $\sigma^2_{\Delta}$ -- $p^{-1}$ relation is $+0.98$, and
the weighted least-squares solution for the fit
\begin{equation}
\label{varianceFeH}
\sigma^2_{\Delta}=\left.\sigma^2_0(\Delta)\right/\!p,
\end{equation}
yields $\sigma^2_0(\Delta)=0.0539\pm 0.0075$~dex$^2$, i.e.,
$\sigma_0(\Delta)=0.232\pm 0.016$~dex. (The constant term in
Eq.~(\ref{varianceFeH}) is set to zero to avoid formally negative
values of variance $\sigma^2_{\Delta}$ for large $p$.) The
resulting variance over all common GCs 
$\langle\sigma^2_0(\Delta)\rangle=0.0159\pm 0.0022$~dex$^2$ 
and the average ${\rm [Fe/H]}_{03}$ uncertainty of 0.09~dex
(see Carretta et al.\ 2009) give an estimate of the uncertainty of
${\rm [Fe/H]}_{10}$ of
$\sqrt{\langle\sigma^2_0(\Delta)\rangle-0.09^2}=0.088\pm
0.012$~dex. Thus, the average ${\rm [Fe/H]}$ uncertainties in H03
and H10 may be considered to be the same. We therefore 
adopted the following formulas for the ${\rm [Fe/H]}_{10}$
uncertainty as a function of $p$:
\begin{equation}
\sigma ({\rm [Fe/H]}_{10})= \frac{\sigma_0 ({\rm [Fe/H]}_{10})}{\sqrt{p}},
\label{sigmaFeH}
\end{equation}
\begin{equation}
\sigma_0 ({\rm [Fe/H]}_{10})= \sigma_0(\Delta)/\sqrt{2}=0.164\pm 0.011{\rm\  dex}.
\label{sigma0FeH}
\end{equation}
In particular, $\sigma ({\rm [Fe/H]}_{10})= 0.16,0.12,0.046$~dex for $p=1,2,13$,
respectively. 

The [Fe/H] weights adopted in H10 are well correlated with  [Fe/H] uncertainties and therefore
should be allowed for in modelling the spatial metallicity distribution of GCs. The
$\sigma ({\rm [Fe/H]}_{10})$ values found here can be used as estimates of the absolute values of 
[Fe/H] uncertainty.


\begin{figure}
\center%
\vspace*{-9pt}%
\hspace*{-6pt}%
\includegraphics[angle=0,width=0.5\textwidth,clip]{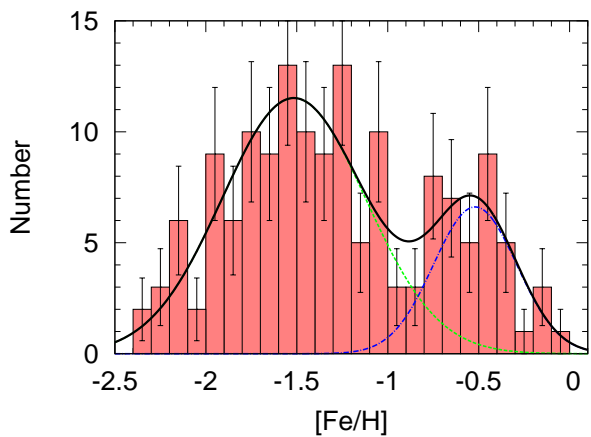}%
\caption%
{Metallicity distribution of GCs for the new H10 scale with the binormal fit.%
}%
\label{fig:DF}
\end{figure}



\begin{figure*}
\center%
\vspace*{-26pt}%
\hspace*{-46pt}%
\includegraphics[clip=true,angle=0,width=0.62\textwidth,clip]{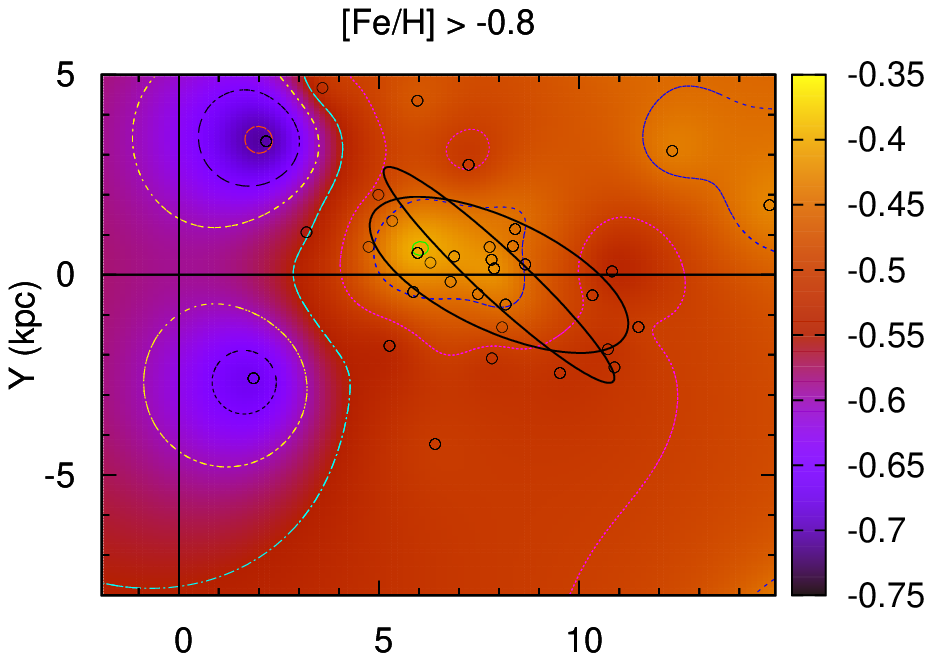}%
\hspace*{-55pt}%
\includegraphics[angle=0,width=0.582\textwidth,clip]{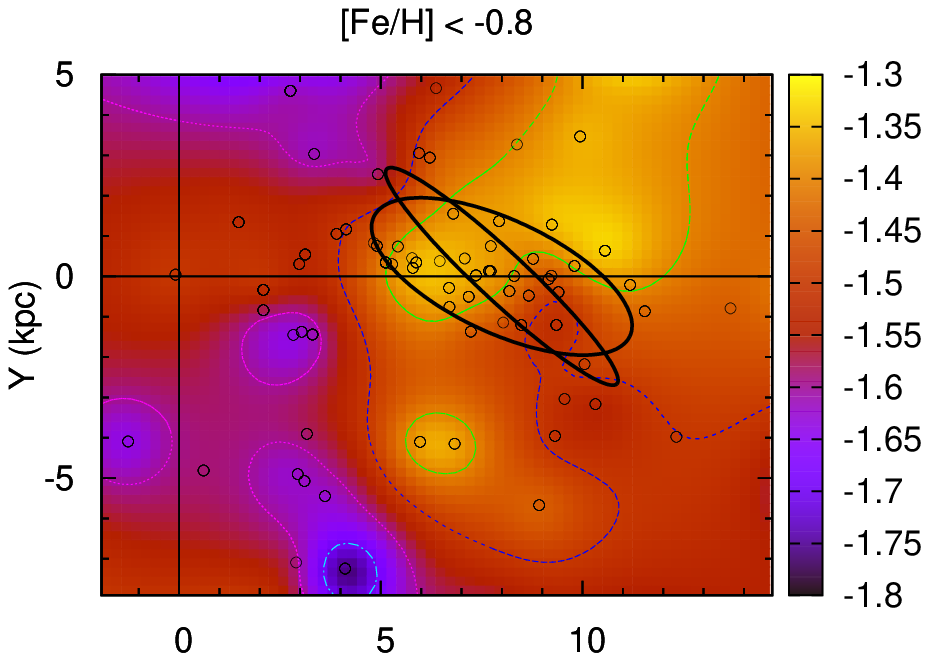}%
\vspace*{-52pt}%

\hspace*{-48pt}%
\raisebox{3pt}{%
\includegraphics[angle=0,width=0.5585\textwidth,clip]{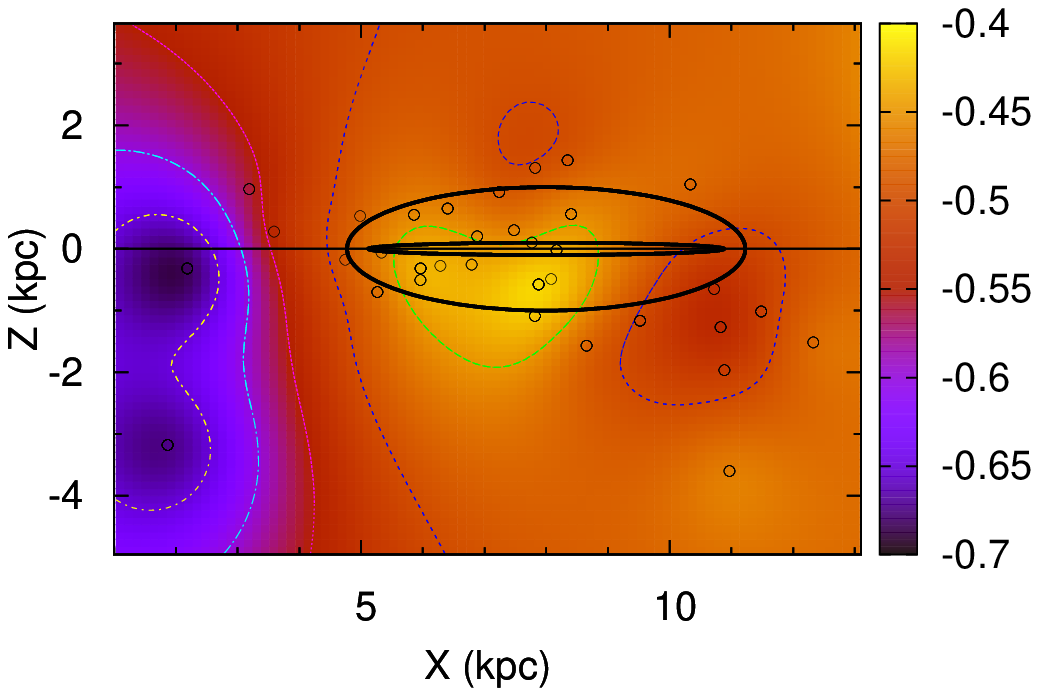}%
}%
\hspace*{-22pt}%
\includegraphics[angle=0,width=0.508\textwidth,clip]{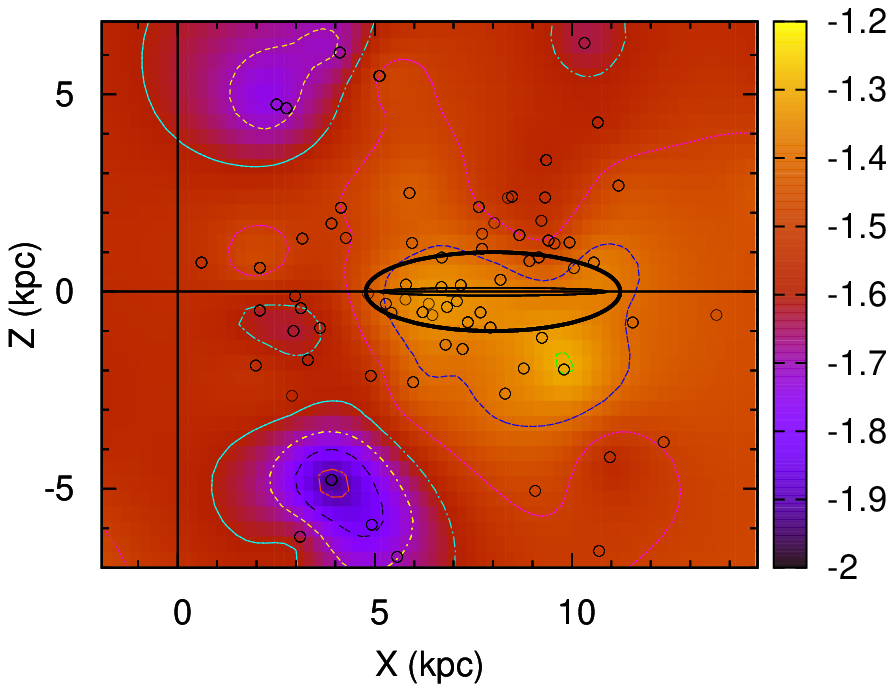}%
\vskip-4mm
\caption%
{Metallicity maps for metal-rich (\emph{left}) and metal-poor (\emph{right}) subsystems of GCs. The
weighted smoothing was performed  with the Cauchy kernel ($d=1$~kpc).  The open circles
represent individual GCs. The thick solid lines outline the Galactic bar ($a=3.5$~kpc,
$b=1.4$~kpc, $c=1.0$~kpc, an angle of~$25^\circ$) and the long bar ($a=3.9$~kpc, $b=0.6$~kpc, $c=0.1$~kpc, an angle
of~$43^\circ$); see Gardner \& Flynn (2010). The Sun is at $(X,Y,Z)=(0,0,0)$, the Galactic
center is at $(X,Y,Z)=(8.0,0,0)$.}
\label{fig:FeHmaps}
\end{figure*}


\newpage

\section{Metallicity distribution and subsystems of~globular clusters}

From this point on, we shall use only the [Fe/H] estimates from H10.

Figure~\ref{fig:DF} shows the metallicity distribution for all 152 GCs in H10. The
figure shows that the binormal model fits the [Fe/H] data well. The maximum-likelihood
fit yields the maxima at $\text{[Fe/H]}=-1.52\pm 0.06$ and $-0.52^{+0.06}_{-0.07}$ with
the standard deviations of $\sigma_{\text{[Fe/H]}}=0.39^{+0.05}_{-0.04}$ and  $0.23^{+0.05}_{-0.04}$
respectively. The metallicity threshold separating metal-pour and metal-rich GCs is found
to be $-0.83^{+0.10}_{-0.11}$. Excluding GCs with low weights ($p=1,2,3$) has no
significant effect on the results.

Based on  our best-fit solutions for the distribution of GC metallicities [Fe/H]  (Fig.~\ref{fig:DF}),
the [Fe/H] versus $R$ and [Fe/H] versus~$Z$ relations  ($Z$ is the distance from the
Galactic plane), we assume that the boundary  between  metal-pour and metal-rich GC
subsystems is at ${\rm [Fe/H]}=-0.8$ in the new  metallicity scale of H10.


Our attempts to directly solve the set of equations
\begin{equation}
\label{gradient}
{\rm [Fe/H]}(R)= f_0+f_1 R,
\end{equation}
\begin{equation}
\label{R}
R= \sqrt{R_0^2 +r^2 \cos ^2b -2R_0 r \cos l \cos b},
\end{equation}
(here $l$ and $b$ are the Galactic coordinates of the GC and $r$ is the
heliocentric distance to the GC) for $R_0$, $f_1$, and
$f_0$ for these two GC subsystems {\em individually\/} failed to
produce well-conditioned results for $R_0$. This has cast doubt
on the correctness of axisymmetric models like~(\ref{gradient})
for the spatial metallicity distribution of GCs.

To analyze the GC metallicity field, i.e., the smoothed dependence of GC metallicity on
spatial coordinates, we produced metallicity maps in the $XY$, $YZ$, $XZ$ planes for
various subsamples of GCs. All maps presented in this paper were obtained by the weighted
smoothing with the Cauchy kernel ($d=1$~kpc). For the metal-rich and metal-poor subsystems
of GCs, the $XY$ metallicity maps, as well as the space distribution in the $XY$ plane,
indicate a centrally concentrated, bar-like configuration with the parameters that agree closely
with those of the Galactic bar (Fig.~\ref{fig:FeHmaps}, top panels). This ``bar
component'' of GCs, which is more pronounced for metal-rich GCs, also shows up for
metal-poor GCs. Note that formal metallicity gradient along the long axis of
bar is present in the $XY$ maps for both GC subsystems. However, as is evident from the $XZ$
maps (Fig.~\ref{fig:FeHmaps}, bottom panels), this gradient is in each case due to the fact
that more metal-rich GCs are always or mostly located within the near side of the bar (in the far
side, GCs are not visible, probably because of high extinction), and a group of the more
metal-poor GCs is  located, judging from their~$|Z|$, near the far side of the bar, but
outside it, and appear to be seen inside the bar only in the $XY$ projection.

\begin{figure}

\vspace*{-10pt}
\includegraphics[angle=0,width=0.47\textwidth,clip]{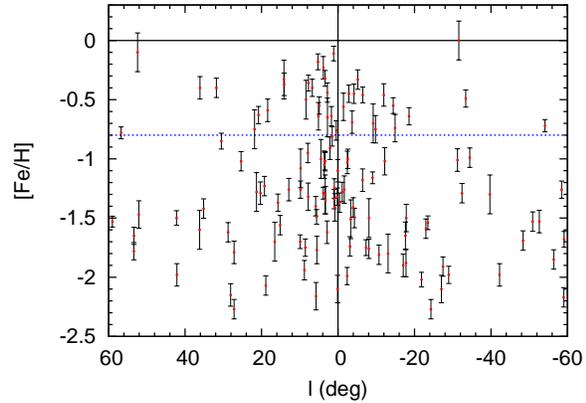}%
\caption%
{Metallicity as a function of Galactic
longitude for globular clusters. Error bars are computed by
Eqs.~(\ref{sigmaFeH})--(\ref{sigma0FeH}). The dotted line marks the metallicity boundary
${\rm [Fe/H]}=-0.8$ between the two subsystems of GCs.}
\label{fig:l-FeH}
\end{figure}

The bar-like configurations in the $XY$ metallicity maps suggest
the presence of inclined elongated structures in the plot of
[Fe/H] versus  Galactic longitude. Figure~\ref{fig:l-FeH} shows that at least
two such structures are actually present, with ${{\rm [Fe/H]}\sim
-1.4}$ and $-0.5$ at ${l=0^\circ}$ for the metal-poor and metal-rich
subsystems of GCs, respectively. For the metal-rich GCs, the
structure is especially long -- it is seen to
extend to
${|l|\sim 30^\circ}$, i.e., it goes beyond the bar boundary. It is not
improbable that more structures exist in the GC system: (i) a
smaller subsystem with ${\rm [Fe/H]}\lesssim -1$ at $l=0^\circ$,
which forms an individual elongated configuration in Fig.~\ref{fig:l-FeH} and,
maybe, shows up in the [Fe/H] distribution (Fig.~2);
(ii)~possible breakdown of metal-rich subsystem into two
groups with parallel chains of points in Fig.~\ref{fig:l-FeH}.


\begin{figure}[ht]
\center%
\vspace*{-37pt}%
\hspace{-25pt}%
\includegraphics[angle=0,width=0.5\textwidth,clip]{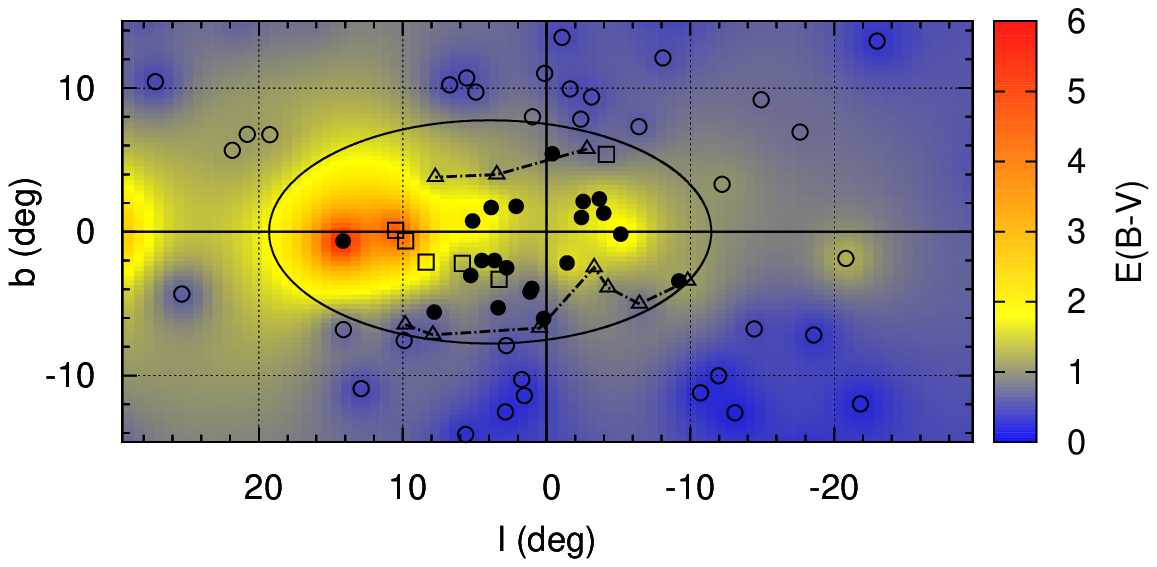}%

\vspace{-50pt}%
\hspace{-25pt}%
\includegraphics[angle=0,width=0.5\textwidth,clip]{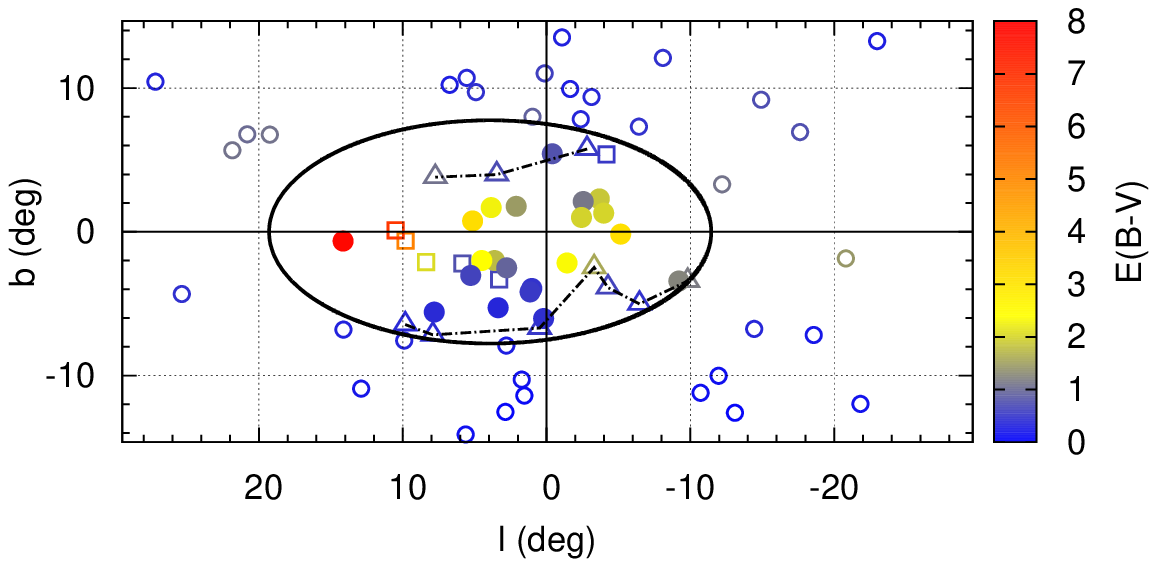}%
\vspace{-20pt}%
\caption%
{\emph{Top}: Foreground reddening map and location of globular clusters projected on the sky in
the Galactic bar region. \emph{Bottom}: The same as in the top panel, but with the 
${E}(B-V)$ values shown for each cluster individually. The solid line shows the
projection of the boundary of the Galactic bar on the sky. The filled circles show the
clusters located inside the Galactic bar. Clusters located outside the bar are plotted as open symbols. The
squares and triangles show the clusters located in the foreground and background relative to
the bar, respectively. The chain-dotted lines connect clusters located behind the bar. The
parameters of the Galactic bar are the same as in Fig.~\ref{fig:FeHmaps}.
}
\label{fig:EB-V:lb}
\end{figure}


\section{Selection effects in the spatial distribution of globular
clusters}

The maps in Fig.~\ref{fig:FeHmaps} suggest that most of GCs at $|Z|\la 1$~kpc are
not detected 
behind the Galactic center. This effect is obviously asymmetric with respect
to the Galactic center and therefore is unlikely to be due to 
dynamic causes. Needless to say,  the deficit of GCs in the ``post-central'' region
of the Galaxy is more likely due to extinction. In this section, we try to verify this 
hypothesis and examine the selection effect in more detail.

Figure~\ref{fig:EB-V:lb} shows the foreground reddening map (top panel), H10 reddening values 
${E}(B-V)$ indicated for each GC individually (bottom panel) along with the
distribution of GCs in projection on the sky for the Galactic bar region. Hereafter we
plot in all figures only the contour of the Galactic bar, because this bar and the long
bar appear to be parts of the same feature (see Athanassoula 2012). The filled circles in Fig.~\ref{fig:EB-V:lb}
and in following figures show the GCs located inside the Galactic bar, and all
open symbols show the GCs located outside the bar. The open squares and triangles
show the GCs located before and behind the bar, respectively.

Figure~\ref{fig:EB-V:lb} demonstrates the existence of a post-central region of avoidance in the system of GCs:
there are no GCs between the chain-dotted lines behind the Galactic bar,
although there are GCs in front and inside the bar. This suggests that the absorbing
matter concentrates not only in the Galactic disk, but also in the Galactic bar. The
reddening map in Fig.~\ref{fig:EB-V:lb} (top panel) is consistent with such speculation.

\begin{figure*}
\center%
\vspace*{-65pt}%
\hspace*{-45pt}%
\includegraphics[angle=0,width=1.07 \textwidth,clip]{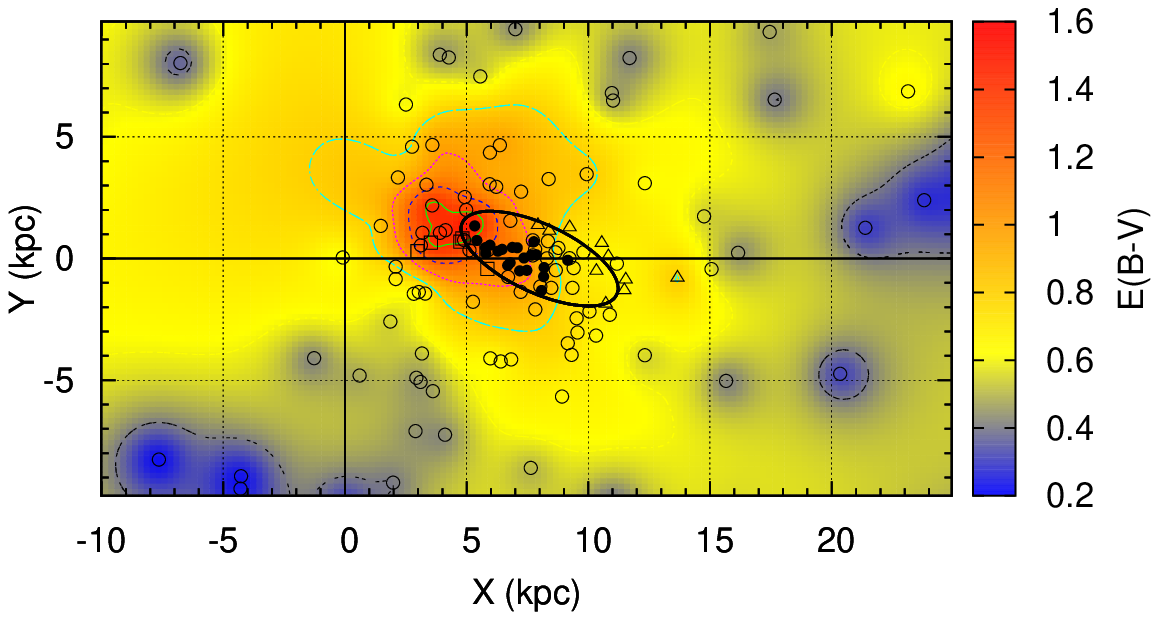}%

\vspace{-60pt}%
\hspace*{-25pt}%
\includegraphics[angle=0,width=0.545 \textwidth,clip]{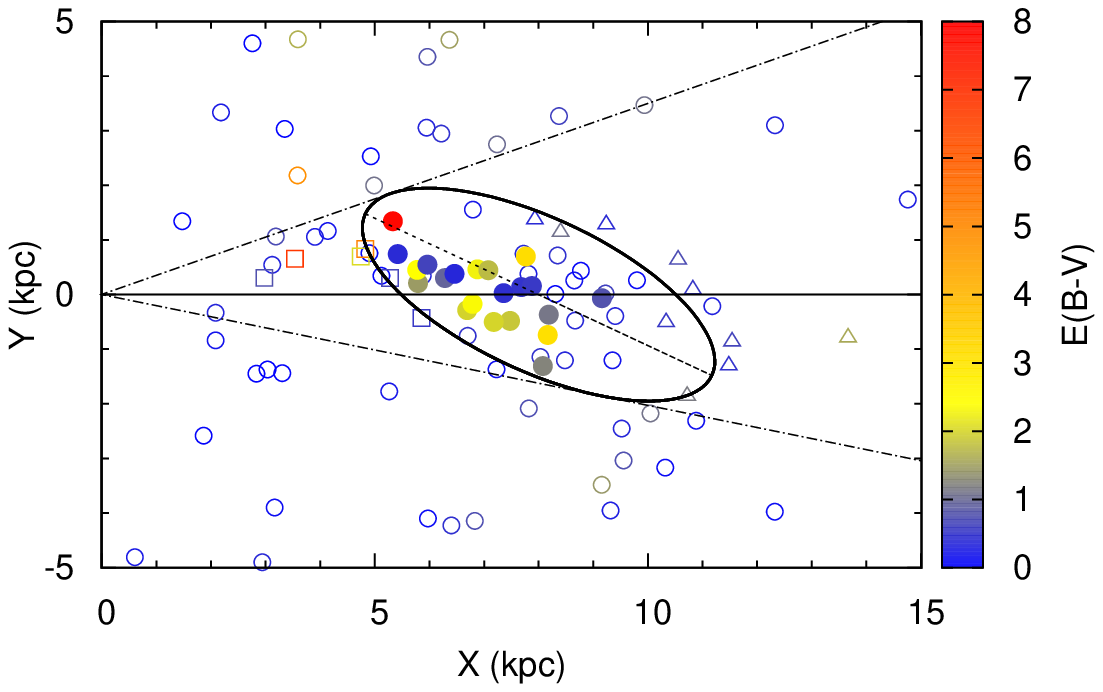}%
\hspace{-19pt}%
\includegraphics[angle=0,width=0.545 \textwidth,clip]{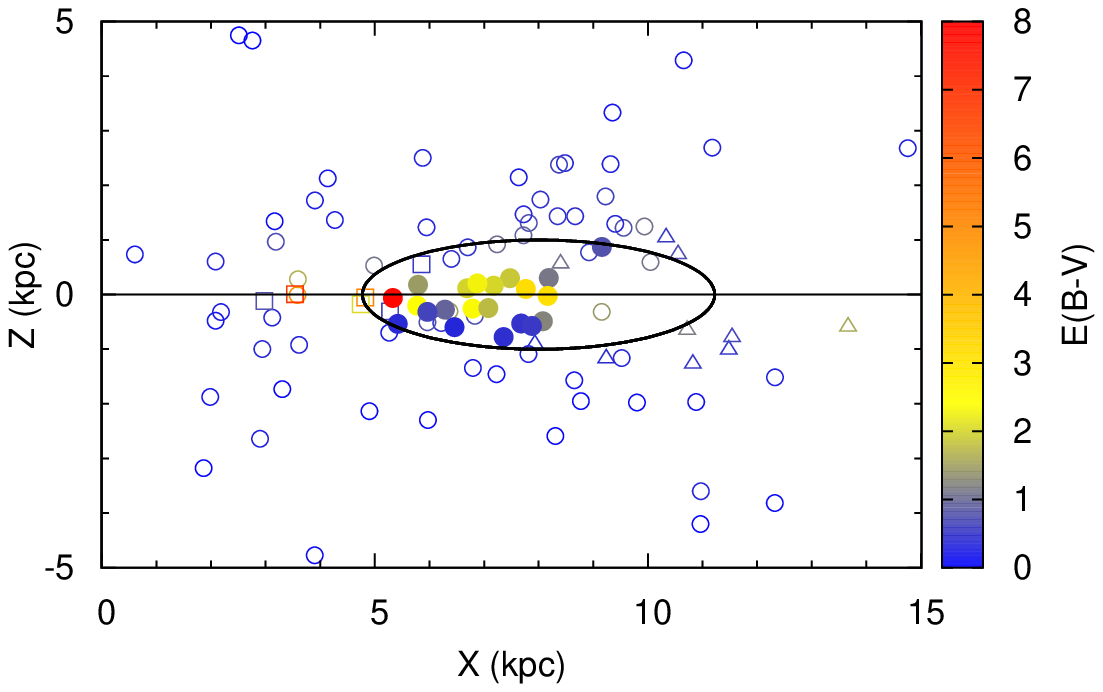}%
\vskip-4mm
\caption%
{\emph{Top}: Foreground reddening map and the location of globular clusters projected on the Galactic
plane. \emph{Bottom left}: Same as in the top panel, but with the reddening values ${E}(B-V)$
shown for each cluster individually. The chain-dotted lines are  tangents to the figure
of the Galactic bar in the Galactic plane and the dashed line shows the major axis of the bar. 
\emph{Bottom right}: Same as in the bottom left panel, but for the projection onto the $XZ$ plane. The
solid line shows the boundary of the projection of the Galactic bar on the corresponding plane. Other
symbols are the same as in Fig.~\ref{fig:EB-V:lb}. \vspace{-2mm} }%
\label{fig:EB-V:xyz}
\end{figure*}

The reddening map and the  spatial distribution of reddening in the $XY$ plane (top and left
bottom panels of Fig.~\ref{fig:EB-V:xyz} respectively) also indicate the existence of
a bar (or at least a central) component of extinction. Moreover, almost all GCs inside 
the bar are distributed within the near side of the bar with a sudden cutoff along the
major axis of the bar in the first Galactic quadrant and along the $X=\mathrm{const}$ line
in the fourth quadrant (the same panels of Fig.~\ref{fig:EB-V:xyz}). NGC~6355 at
$(X,Y,Z)=(9.16,-0.07,0.87)$~kpc is not an exception to this rule: it is located almost
exactly at the boundary of the bar (Fig.~\ref{fig:EB-V:xyz}, right bottom panel) is just
projected onto the empty region.


We thus conclude that all these results can be explained only by the
existence of a bar extinction component which produces a sudden GCs' distribution cutoff
in the directions with the strongest extinction. Note that this conclusion does not depend
on the adopted parameters of the bar, because the existence of the post-central region of
avoidance found does not depend on these parameters.

\begin{figure}
\center%
\vspace*{-20pt}
\hspace{-32pt}%
\includegraphics[angle=0,width=0.5 \textwidth,clip]{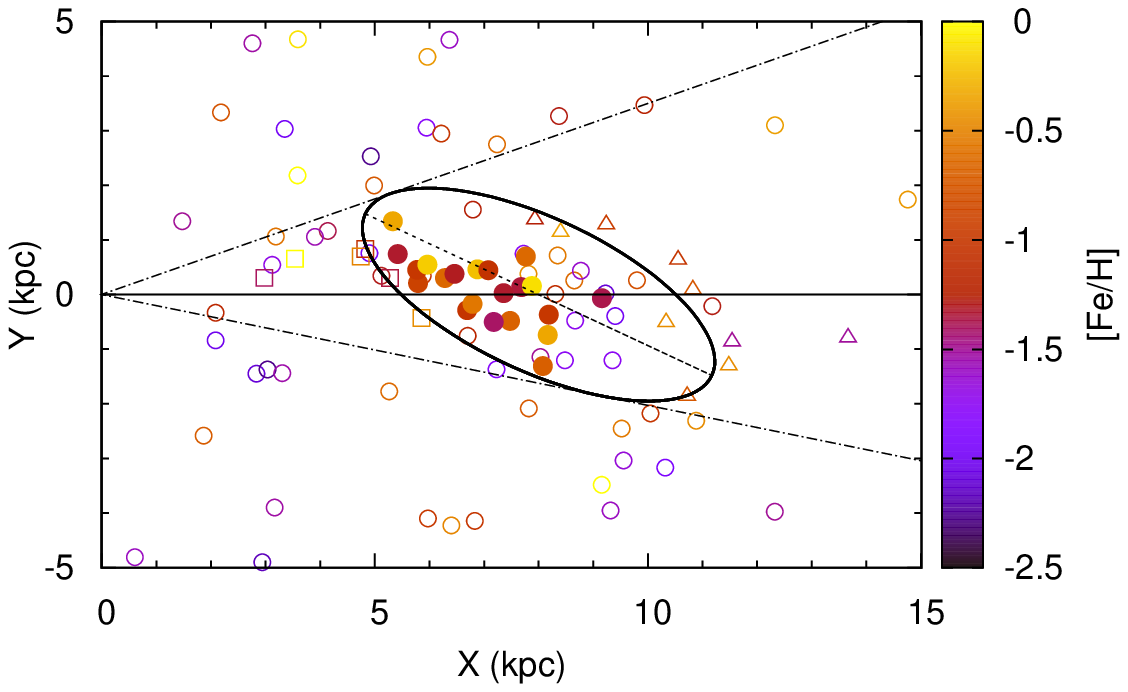}%

\vspace{-35pt}%
\hspace{-32pt}%
\includegraphics[angle=0,width=0.5 \textwidth,clip]{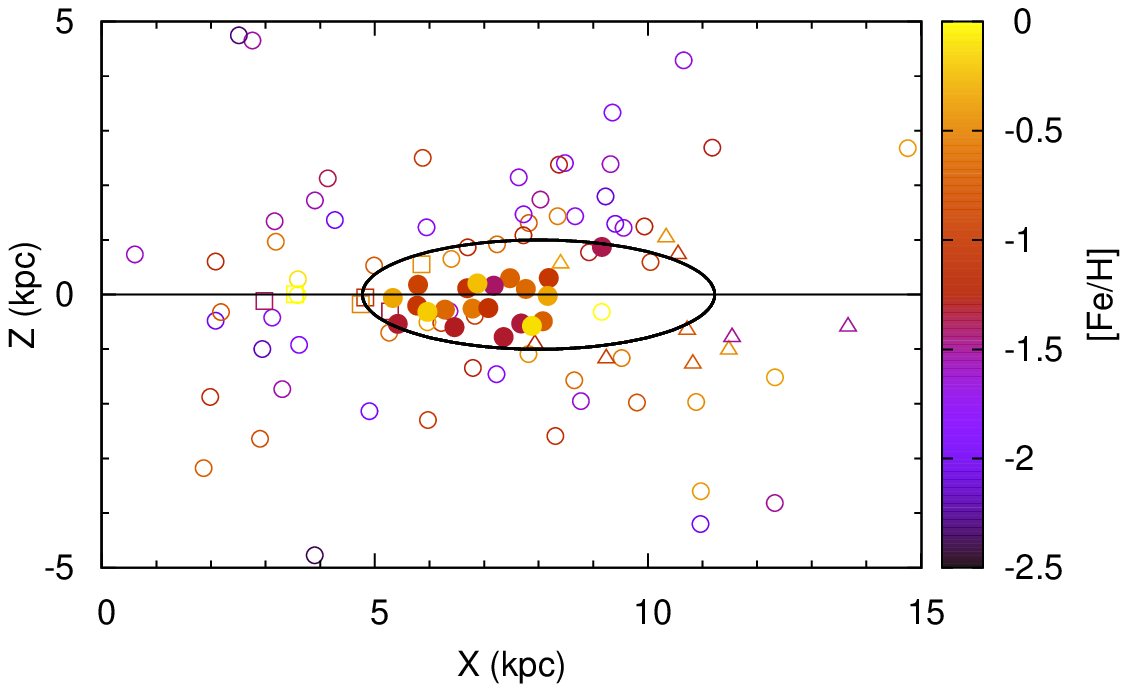}%
\vspace{-15pt}%
\caption%
{The same as in bottom panels of Fig.~\ref{fig:EB-V:xyz}, but for the metallicity of
globular clusters.
}%
\label{fig:FeH:xyz}

\end{figure}

The distribution of GCs in the $XY$ and $XZ$ planes with individual metallicities $\mathrm{[Fe/H]}$
indicated for each cluster  (Fig.~\ref{fig:FeH:xyz}) illustrates how the
presence of a bar component within both GC subsystems combined with selection
due to the concentration of absorbing matter in the Galactic bar produce in
metallicity maps (Fig.~\ref{fig:FeHmaps})  structures associated with the bar.


\begin{figure}
\center%
\vspace*{-5pt}
\hspace*{-6pt}%
\includegraphics[angle=0,width=0.5\textwidth,clip]{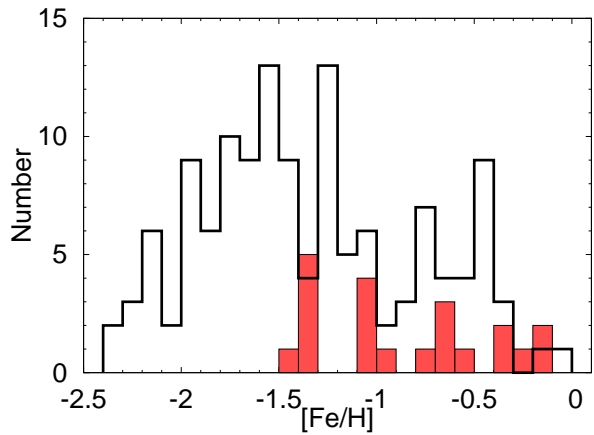}%
\caption%
{A comparison between the metallicity distributions of GCs located inside (solid boxes) and
outside (steps) the Galactic bar.
}%
\label{fig:DF:in-out}
\end{figure}

The bar GCs differ noticeably from other GCs in terms of iron abundances (Figs.~\ref{fig:FeH:xyz} and
\ref{fig:DF:in-out}). Figure~\ref{fig:DF:in-out} shows that in the metal-rich subsystem 
the fraction of GCs with the highest abundances is greater among the bar GCs than among
GCs located outside the bar; moreover, the bar component of metal-poor subsystems contains only GCs
with $\mathrm{[Fe/H]}>-1.50$.

\section{Discussion}

The effect of bar-like  configurations on GC metallicity maps suggests that the bar GCs
formed within the already existing Galactic bar or were later locked in
resonance with the bar. In the first case, the implication is  that the Galactic bar may
have the age of 10~Gyr or more. The presence of parallel elongated structures in the [Fe/H] versus $l$
plot (Fig.~\ref{fig:l-FeH}) is rather indicative of bar-induced resonance effects. In
any case, these GCs seem to be associated with the Galactic bar.

Note  that previously Burkert \& Smith (1997) also identified the subsystem of bar
clusters, but only among the metal-rich GCs, based on an analysis of the
space distribution and kinematics. It is unlikely that this is due to chance.

Regardless of the details of the origin of  this effect, it is clear
that the spatial distribution of GC metallicities is not
axisymmetric. Hence justified and strong constraints on $R_0$ can be obtained
only in terms of a non-axisymmetric model for this distribution and 
with the allowance for the fact that the GC population consists of
several subsystems. The sizes of bar-like configurations and the number of GCs 
located inside them lead us to expect statistical uncertainties of 
$0.4$--$0.5$~kpc for $R_0$ estimates based on the metallicity data for each of two main subsystems of
GCs. Hence we conclude that this approach appears to be
promising.


Observational incompleteness of GCs in the far side of the Galactic bar and in the
post-central region shows that the allowance for the selection effect due solely to extinction
in the layer of constant scale height (e.g., Racine \& Harris 1989; Surdin 1999) seems
to be insufficient to eliminate the corresponding systematic errors in  $R_0$ estimates. 
This may  explain why the GC-based $R_0$ estimates are systematically smaller
than the best  $R_0$ values (Table~\ref{tab:R0average}).

\acknowledgements
This study was partly supported by the Russian President Grant for the Support
of the Leading Scientific Schools of Russia (No.~NSh-3290.2010.2).

Russian President Grant for Support
to Leading Scientific School of Russia

\end{document}